\documentclass[aps,prd,superscriptaddress,amssymb,eqsecnum,showpacs,showkeyes,secnumarabic,graphics,floatfix,nofootinbib,tightenlines,longbibliography]{revtex4-1}

\usepackage{graphicx}
\usepackage{epstopdf}
\usepackage{epsfig}
\usepackage{subcaption}
\usepackage{bm}
\usepackage{cancel}
\usepackage{dcolumn}
\usepackage{amsmath}
\usepackage{mathtools}
\usepackage{hhline}
\usepackage[utf8]{inputenc}
\usepackage{multirow}

\begin{document}

\newcommand{\newc}{\newcommand}

\newc{\be}{\begin{equation}}
\newc{\ee}{\end{equation}}
\newc{\ba}{\begin{eqnarray}}
\newc{\ea}{\end{eqnarray}}
\newc{\bea}{\begin{eqnarray*}}
\newc{\eea}{\end{eqnarray*}}
\newc{\D}{\partial}
\newc{\ie}{{\it i.e.} }
\newc{\eg}{{\it e.g.} }
\newc{\etc}{{\it etc.} }
\newc{\etal}{{\it et al.}}
\newc{\lcdm}{$\Lambda$CDM }
\newcommand{\nn}{\nonumber}
\newc{\ra}{\Rightarrow}

\begin{flushright}
{{\today}\vspace{1.5cm}} 
\end{flushright}

\title{Sudden Future Singularities and  their observational signatures in Modified Gravity\footnote{Contribution to the School and Workshops on Elementary Particle Physics and Gravity, 2-28 September 2017, Corfu, Greece; to appear in the Proceedings of Science.}\vspace{1.8cm}}

\author{{\bf Andreas Lymperis}\vspace{0.5cm}}
\email{alymperis@upatras.gr} 
\affiliation{Department of Physics, University of Patras, 26500 Patras, Greece \vspace{2cm}}

\begin{abstract}
\section*{abstract} 
We verify the existence of Generalized Sudden Future
  Singularities (GSFS) in quintessence models with scalar field
  potential of the form $V(\phi)\sim \vert \phi\vert^n$ where $0<n<1$
and in the presence of a perfect fluid,
  both numerically and analytically, using a proper generalized expansion ansatz for the
  scale factor and the scalar field close to the singularity. This
  generalized ansatz includes linear and quadratic terms, which
  dominate close to the singularity and cannot be ignored when
  estimating the Hubble parameter and the scalar field energy
  density; as a result, they are important for analysing the
  observational signatures of such singularities. We derive analytical
  expressions for the power (strength) of the singularity in terms of
  the power $n$ of the scalar field potential. We then extend the
  analysis to the case of scalar tensor quintessence models with the
  same scalar field potential in the presence of a perfect fluid, and
  show that a Sudden Future Singularity (SFS) occurs in this case. We derive both analytically and numerically the strength of the singularity in terms of the power $n$ of the scalar field potential.
\end{abstract}

\maketitle

\section{Introduction}

Latest evidence of an accelerating Universe \cite{1, 2, 3, 4, 5, 6},
has opened new windows in the context of the study of physics in
cosmological scales, and has lead to the consideration of models alternative to \lcdm. Such models include modifications of GR (modified Gravity) \cite{7, 8}, scalar field dark energy (quintessence) \cite{9, 10}, physically motivated forms of fluids \eg Chaplygin gas \cite{11, 12} etc.

Some of these dark energy models predict the existence of exotic cosmological singularities, involving divergences of the scalar spacetime curvature and/or its derivatives. These singularities can be either  geodesically complete \cite{13, 14, 15, 16} (geodesics continue beyond the singularity and the Universe may remain in existence) or geodesically incomplete \cite{17, 18} (geodesics do not continue beyond the singularity and the Universe ends at the classical level). They appear in various physical theories such as superstrings \cite{19}, scalar field quintessence with negative potentials \cite{20}, modified gravities and others \cite{21, 22}.

The divergence of the scale factor and/or its derivatives leads to divergence of scalar quantities like the Ricci scalar, thus to different types of singularities or `cosmological milestones' \cite{23, 25, 26}. However geodesics do not necessarily end at these singularities and if the scale factor remains finite, they are extended beyond these events \cite{22} even though a diverging impulse may lead to dissociation of all bound systems in the Universe at the time $t_s$ of these events\cite{24}.  

Thus, singularities can be classified \cite{27} according to the behaviour of the scale factor $a(t)$, and/or its derivatives at the time $t_s$ of the event or equivalently, and the energy density and pressure of the content of the universe at the time $t_s$. A classification of such singularities and their properties is shown in Table \ref{TabI}.

\begin{table}[h]
\caption{Classification and properties of cosmological singularities.}\label{TabI}
\resizebox{1\textwidth}{!}
{
\begin{tabular}{c c c c c c c c c c} \\
  \hline
 Name & $t_{sing}$ & $a(t_{s})$ & $\rho(t_{s})$ & $p(t_{s})$ & $\dot p(t_{s})$ & $w(t_{s})$ & T & K & Geodesically \\ 
 \hline\hline
 Big-Bang (BB) & 0 & 0 & $\infty$ & $\infty$ & $\infty$ & finite & strong & strong &  incomplete \\ 
 \hline
 Big-Rip (BR) & $t_{s}$ & $\infty$ & $\infty$ & $\infty$ & $\infty$ & finite & strong & strong & incomplete \\
 \hline
 Big-Crunch (BC) & $t_{s}$ & 0 & $\infty$  & $\infty$ & $\infty$ & finite & strong & strong &  incomplete \\
 \hline
 Little-Rip (LR) & $\infty$ & $\infty$ & $\infty$ & $\infty$ & $\infty$ & finite & strong & strong & incomplete \\
 \hline
 Pseudo-Rip (PR) & $\infty$ & $\infty$ & finite & finite & finite & finite & weak & weak & incomplete \\
 \hline
 Sudden Future (SFS) & $t_{s}$ & $a_{s}$ & $\rho_{s}$ & $\infty$ & $\infty$ & finite & weak & weak & complete \\  
 \hline
 Big-Brake (BBS) & $t_{s}$ & $a_{s}$ & 0 & $\infty$ & $\infty$ & finite & weak & weak & complete \\  
 \hline
 Finite Sudden Future (FSF) & $t_{s}$ & $a_{s}$ & $\infty$ & $\infty$ & $\infty$ & finite & weak & strong & complete \\
 \hline
Generalized Sudden Future (GSFS) & $t_{s}$ & $a_{s}$ & $\rho_{s}$ & $p_{s}$ & $\infty$ &finite & weak & strong & complete \\
 \hline
 Big-Separation (BS) & $t_{s}$ & $a_{s}$ & 0 & 0 & $\infty$ & $\infty$ & weak & weak & complete \\
 \hline
 w-singularity (w) & $t_{s}$ & $a_{s}$ & 0 & 0 & 0 & $\infty$ & weak & weak & complete \\
 \hline
\end{tabular}
}
\end{table}

A particularly interesting type of singularities are the Sudden Future Singularities \cite{21}, which involve violation of the dominant energy condition $\rho \geq |p|$, and divergence of the cosmic pressure of the Ricci Scalar and of the second time derivative of the cosmic scale factor Table \ref{TabI}. The scale factor can be parametrized as 

 \be \label{sfab}
a(t)=\left (\frac{t}{t_{s}} \right )^{m} (a_{s}-1)+1-\left (1-\frac{t}{t_{s}} \right )^{q},
\ee

\noindent where $a_{s}$ is the scale factor at the time $t_{s}$ and $1<q<2$. For this range of the parameter $q$, the scale factor and its first derivative, \ie $a, \dot a$ respectively, and $\rho$ remain finite at $t_{s}$. However,  the quantities $p, \dot \rho$ and $\ddot a$ become infinite. Thus, when the first derivative of the scale factor is finite at the singularity, but the second derivative diverges (SFS singularity \cite{21, 28}), the energy density is finite but the pressure diverges.

In the following, we focus on the quintessence models with a perfect fluid, and investigate the strength of the GSFS both analytically and numerically. We extend the analysis to the case of scalar-tensor quintessence and investigate the modification of the strength of the singularity both analytically (using a proper expansion ansatz) and numerically, by explicitly solving the dynamical cosmological equations.

\section{The setup}

In FRW spacetime with metric

\be
ds^2=-dt^2+a^{2}(t)\bigg[\frac{dr^2}{1-kr^2}+r^2(d\theta^2+\sin^{2}\theta d\phi^2)\bigg]
\ee

\noindent the most general action involving gravity, nonminimally coupled with a scalar field $\phi$, and a perfect fluid is 

\be \label{staction}
S=\int \left [\frac{1}{2}F(\phi)R+\frac{1}{2}g^{\mu \nu} \phi_{;\mu}  \phi_{;\nu}-V(\phi)+\mathcal{L}_{(fluid)} \right ]\sqrt{-g}d^{4}x.
\ee

\noindent where $F(\phi)$ is the nonminimal coupling of gravity to the
scalar field and $\mathcal{L}_{(fluid)}$ the fluid term. We have set
$8\pi G=c=1$ and assume spatial flatness ($k = 0$). In the case of the
scalar-tensor models, corresponding to the action (\ref{staction}), we
assume a non-minimal coupling linear in the scalar field
$F(\phi)=1-\lambda \phi$, even though the results on the type of the
singularity in this class of models
are unaffected by the particular choice of the non-minimal coupling.  

\noindent In the special case where the non-minimal coupling $F(\phi)=1$, the action (\ref{staction}) reduces to the simple case of quintessece models with a perfect fluid

\be \label{qaction}
S=\int \left [\frac{1}{2}R+\frac{1}{2}g^{\mu \nu} \phi_{;\mu}  \phi_{;\nu}-V(\phi)+\mathcal{L}_{(fluid)} \right ]\sqrt{-g}d^{4}x.
\ee

\begin{figure}[!h]
\centering
\vspace{0.3cm}\rotatebox{0}{\vspace{0cm}\hspace{0cm}{\includegraphics{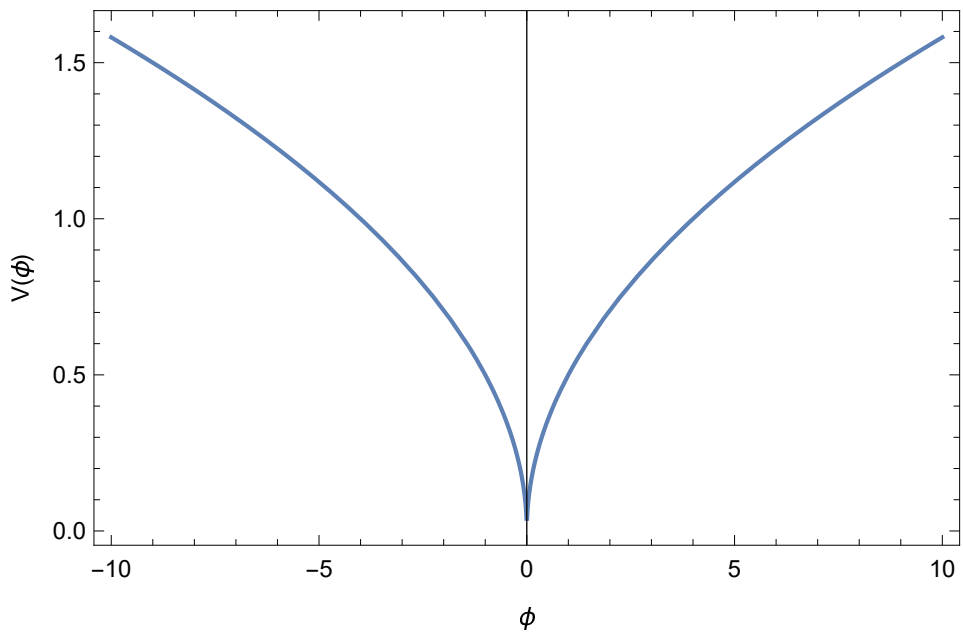}}}
\caption{Dynamical evolution of the scalar field potential $V(\phi)=A|\phi|^{n}$}
\label{fig:fig1}
\end{figure}

The potential $V(\phi)$ is of the form

\be \label{potential}
V(\phi)=A|\phi|^{n},\ \ \ \ \  A>0,
\ee

\noindent with $0<n<1$ and $A$ a constant parameter. The dynamical evolution of the scalar field due to the potential is shown in Fig. \ref{fig:fig1}

It was shown, through a qualitative analysis \cite{30}, that the power law scalar potential (\ref{potential}) leads to singularities at any scale factor derivative order larger than three, depending on the value of the power $n$. In particular, for $k<n<k+1$, with $k>0$, the $(k+2)^{th}$ derivative of the scale factor diverges at the singularity. This is in fact the simplest extension of \lcdm with geodesically complete cosmic singularities and occurs at the time $t_s$, when the scalar field becomes zero ($\phi=0$).

\section{The Quintessence case}

The action in this class of models, is of the form (\ref{qaction}). The energy density and pressure of the scalar field $\phi$, may be written as 

\be \label{rphipphi}
\rho_{\phi}=\frac{1}{2}\dot \phi ^{2}+V(\phi) \ \ \ \ \ \ \ \ and \ \ \ \ \ \ \ \ p_{\phi}=\frac{1}{2}\dot \phi ^{2}-V(\phi).
\ee \

\noindent and we assume that the perfect fluid is pressureless ($p_{m}=0$).

Variation of the action (\ref{qaction}) leads to the dynamical equations 

\be \label{qde1}
3H^2=\frac{3\Omega_{0,m}}{a^{3}}+ \frac{1}{2}\dot{\phi}^2+V(\phi)
\ee

\be \label{qde2}
\ddot{\phi}=-3H\dot{\phi}-An|\phi|^{n-1}  \Theta(\phi)
\ee

\be \label{qde3}
2\dot H=-\frac{3\Omega_{0,m}}{a^{3}}-\dot \phi^2
\ee

\noindent where $a$ is the scale factor, $H=\frac{\dot a}{a}$ is the Hubble parameter, $\rho_{m}=\frac{\rho_{0,m}}{a^{3}}=\frac{3\Omega_{0,m}}{a^{3}}$, $\Omega_{0,m}=0.3$ and 

\be 
\Theta(\phi)=
\begin{cases}
1, & \phi>0 \\
-1, & \phi<0
\end{cases}
\ee. 

From eqs (\ref{qde1}), (\ref{qde3}), it follows that when $t\to t_{s}$ \ie $\phi \to 0$, the Hubble parameter $H$ and its first derivative $\dot H$ remain finite and so does $\dot \phi$. But in eq. (\ref{qde2}) there is a divergence of the term $\phi^{n-1}$ for $0<n<1$ and thus $\ddot \phi \to \infty$ as $\phi \to 0$. $\ddot H$ also diverges at this point due to the divergence of $\ddot \phi$, as follows by differentiating eq. (\ref{qde3}). This implies that the third derivative of the scale factor diverges, and a GSFS occurs at this point (\ie $a_{s}, \rho_{s}, p_{s}$ remain finite but $\dot p \to \infty)$. Thus, the constraints on the power exponents $q,r$ of the diverging terms in the expansion of the scale factor ($\sim (t_{s}-t)^q$ ) and of the scalar field ($\sim (t_{s}-t)^r$ ) are $2<q<3$ and $1<r<2$ respectively (see eqs (\ref{sfa}), (\ref{sfi}) below). It has been shown in \cite{31} that by choosing $q$ to lie in the intervals $(N, N+1)$ for $N\geq 2$, where $N\in \mathbb{Z}^+$, a finite-time singularity occurs in which

\be
\frac{d^{N+1}a}{dt^{N+1}}\to \infty
\ee
 
\noindent but
 
\be
\frac{d^{s}a}{dt^{s}}\to 0, \ \  for \ \  s\leq N\in \mathbb{Z}^+
\ee

\noindent This allows for pressure singularities which are accompanied by divergence of higher time derivatives of the scale factor (divergence of the fourth-order derivative of the scale factor \cite{31} when $p\to \infty$), in Friedmann solutions of higher-order gravity $(f(R))$ theories \cite{32}.

The above qualitative analysis can be extended to a quantitative level by introducing a new ansatz for the scale factor and the scalar field, containing linear and quadratic terms of $(t_{s}-t)$. These terms play an important role, since they dominate in the first and second derivative of the scale factor as the singularity is approached.  

The new ansatz for the scale factor which generalizes (\ref{sfab}), by introducing linear and quadratic terms in $(t_{s}-t)$, is of the form \cite{29}

\be \label{sfa}
a(t)=1+(a_s-1)\left(\frac{t}{t_s}\right)^m+b(t_{s}-t) + c(t_{s}-t)^2 +d(t_{s}-t)^q,
\ee

\noindent where $m=\frac{2}{3(1+w)}$, $w$ the state parameter, $b, c, d$ are real constants to be determined, and $2<q<3$ so that $\dddot a$ diverges at the GSFS. 

\noindent The corresponding expansion of the scalar field $\phi(t)$ in the vicinity of the singularity is of the form

\be \label{sfi}
\phi(t)=f(t_{s}-t)+h(t_{s}-t)^{r}
\ee

\noindent where $1<r<2$ so that $\ddot \phi$ diverges at the singularity and $f,h$ are real constants to be determined.

\begin{figure}[!h]
\centering
\vspace{0.3cm}\rotatebox{0}{\vspace{0cm}\hspace{0cm}{\includegraphics{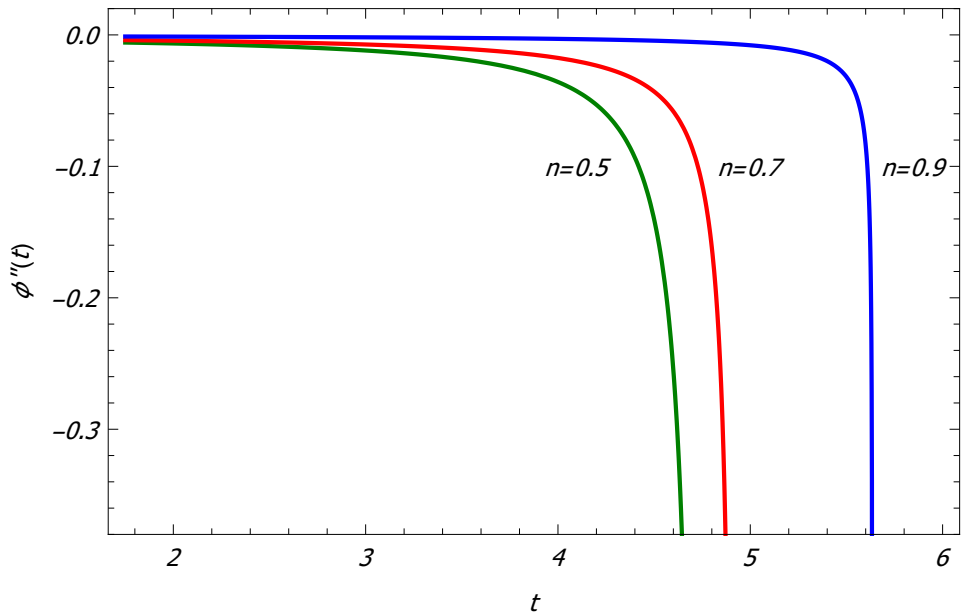}}}
\caption{Numerical solutions of the second time derivative of the scalar field for $n=0.5, 0.7, 0.9$. Notice the divergence at the time of the singularity when the scalar field vanishes.}
\label{fig:fig2}
\end{figure}

From eq. (\ref{qde2}) and differentiated eq. (\ref{qde3}), using the forms of the scale factor (\ref{sfa}) and the scalar field (\ref{sfi}), we get two equations that contain only dominant terms in $(t_{s}-t)$, in which both the left and right-hand sides diverge at the singularity for $0<n<1$, $2<q<3$ and $1<r<2$. Equating the power laws $q$ and $r$ of the divergent terms we obtain

\be \label{qr}
r=n+1
\ee

\be \label{qq}
q=r+1.
\ee

and it follows that

\be \label{qqn}
q=n+2.
\ee
 
\begin{figure}[!h]
\centering
\begin{subfigure}{.5\textwidth}
  \includegraphics[width=.95\linewidth]{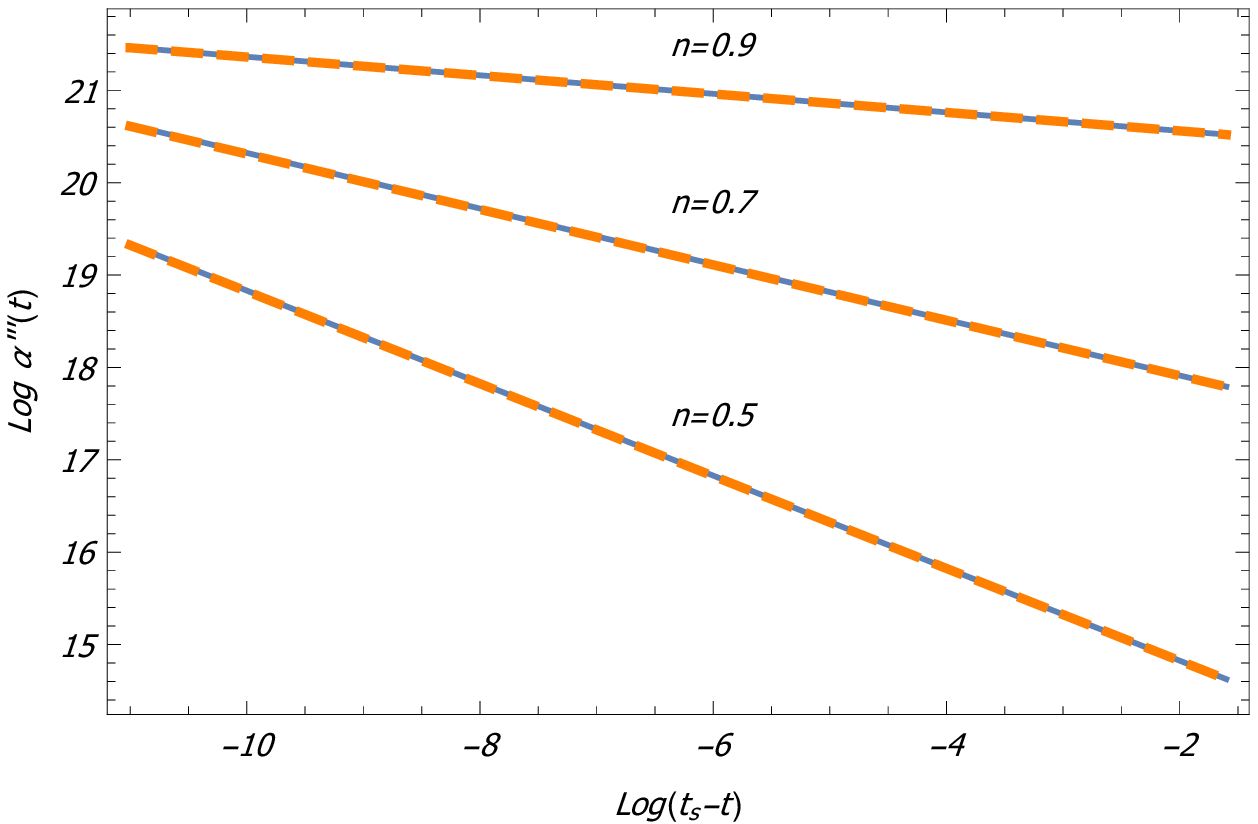}
  \caption{3a}
  \label{fig:fig3}
\end{subfigure}%
\begin{subfigure}{.5\textwidth}
  \centering
  \includegraphics[width=.95\linewidth]{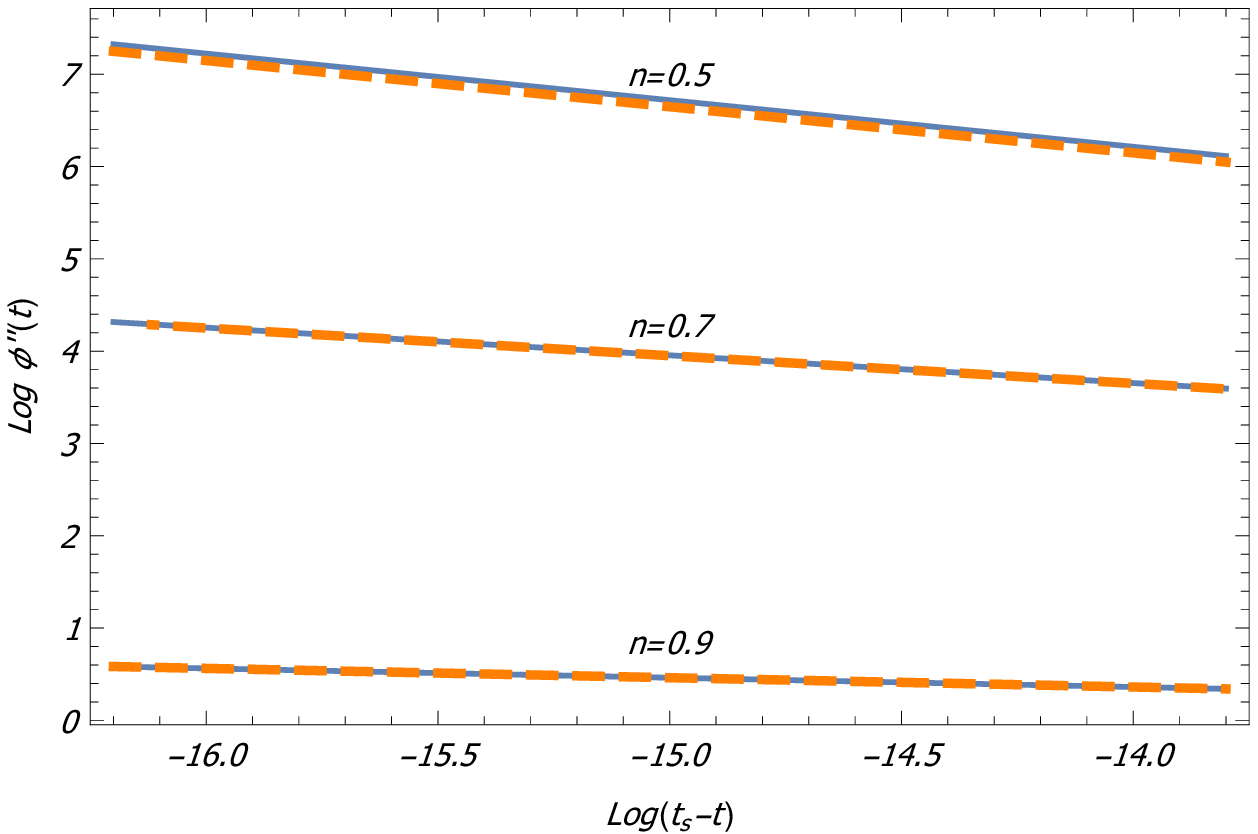}
  \caption{3b}
  \label{fig:fig4}
\end{subfigure}
\caption{Plots of numerical verification of the $q$-exponent (3a) and $r$-exponent (3b) for 3 values of $n$ ($n=0.5, n=0.7$ and $n=0.9$). The orange dashed line, denotes the analytical, while the blue line denotes the numerical solution. As expected the slopes for each n for both $q$ and $r$ are identical.}
\end{figure}

Figure \ref{fig:fig2} shows the divergence of the second derivative of
the scalar field at the time of the singularity. In figures
\ref{fig:fig3}, \ref{fig:fig4} we plot the numerically verified derived power law dependence (eqs (\ref{qr}), (\ref{qqn})) of the scalar field and the scale factor respectively, as the singularity is approached. It is clear that eqs (\ref{qr}), (\ref{qqn}) are consistent with the qualitatively expected range of $r,q$, for $0<n<1$. 

The additional linear and quadratic terms in $(t_{s}-t)$, in the expression of the scale factor (\ref{sfa}), play an important role in the estimation of the Hubble parameter and its derivative as the singularity is aproached. An interesting result arises from the derivation of the relation between the coefficients $b, c$. The relations between these coefficients can lead to relations between the Hubble parameter and its derivative close to the singularity, which in turn correspond to observational predictions, that may be used to identify the presence of these singularities in angular diameter of luminosity distance data. The relation between $b, c$  is of the form  

\be \label{qc}
c=\frac{\rho_{0,m}}{4a^{2}_{s}}-\frac{1}{2}(a_{s}-1)m(m-1)-\frac{[(a_{s}-1)m-b]^{2}}{a_{s}},
\ee

\noindent and thus

\be \label{qdotH}
\dot H=\frac{3\Omega_{0,m}}{2a^{3}_{s}}-3H^{2}
\ee

\noindent and as a function of redshift parameter $z$ at present time

\be \label{qHz}  
H^{2}(z)= \Omega_{0,m}(1+z)^{3}[1-(1+z)^{3}(1+z_{0})^{-3}]+(1+z)^{6}(1+z_{0})^{-6}H^{2}_{0},
\ee

\noindent where $H_{0}, z_{0}$ are the Hubble and redshift parameter respectively at present time. This result may be used as observational signature of such singularities in this class of models.

In the absence of the perfect fluid, the strength of the singularity
remains unaffected. This means that the evaluated relations of $r$ and
$q$ (eqs (\ref{qr}), (\ref{qqn})) respectively, are exactly the same. The Hubble parameter and its derivative in this case is

\be \label{qdotHwm}
\dot H=-3H^2
\ee

\noindent and as a function of redshift parameter $z$ at present time

\be \label{qHzwm}
H(z)=\frac{H_{0}(1+z)^{3}}{(1+z_{0})^3}.
\ee

\noindent These are the reduced relations of eqs (\ref{qdotH}) and (\ref{qHz}) respectively, for $\rho_{0,m}=0$.

\section{Modified Gravity: The Scalar-Tensor Quintessence case}

The action of the theory, in this class of models, is of the form (\ref{staction}). The corresponding dynamical equations are 

\be \label{stde1}
3FH^{2}=\frac{3\Omega_{0,m}}{a^{3}}+\frac{\dot \phi^2}{2}+V-3H\dot{F}
\ee

\be \label{stde2}
\ddot \phi+3H\dot \phi-3F_{\phi}\bigg(\frac{\ddot a}{a}+H^2\bigg)+An|\phi|^{(n-1)} \Theta(\phi)=0
\ee

\be \label{stde3}
-2F\bigg(\frac{\ddot a}{a}-H^{2}\bigg)=\frac{3\Omega_{0,m}}{a^{3}}+\dot \phi^{2}+\ddot F-H\dot F,
\ee

\noindent where $F_{\phi}=\frac{dF}{d\phi}$. From eq. (\ref{stde1}), it is clear that $H, \dot \phi, F, \dot F$ all remain finite when $\phi \to 0$ ($t\to t_{s}$). However, in eq. (\ref{stde2}) there is a divergence of the term $V_{\phi}$ for $0<n<1$ and $\ddot \phi \to \infty$ as $\phi \to 0$. This means that $\ddot F\to \infty$ because of the generation of the second derivative of $\phi$ that leads to a divergence of $\ddot a$ in eq. (\ref{stde3}). Clearly, an SFS singularity (Table \ref{TabI}) is expected to occur in scalar-tensor quintessence models, as opposed to the GSFS singularity in the corresponding quintessence models. Thus, the constraints on the power exponents $q,r$ in this case are $1<q<2$ and $1<r<2$ respectively.

From the above dynamical equations, using the same parametrizations
(\ref{sfa}), (\ref{sfi}) for the scale factor and the scalar field respectively and keeping only the dominant terms, the values for $r$ and $q$ are

\be \label{stq}
q=r
\ee

\be \label{str}
r=n+1,
\ee

\noindent which  leads to

\be \label{stqn}
q=n+1.
\ee

\begin{figure}[!h]
\centering
\begin{subfigure}{0.5\textwidth}
\includegraphics[width=0.95\linewidth]{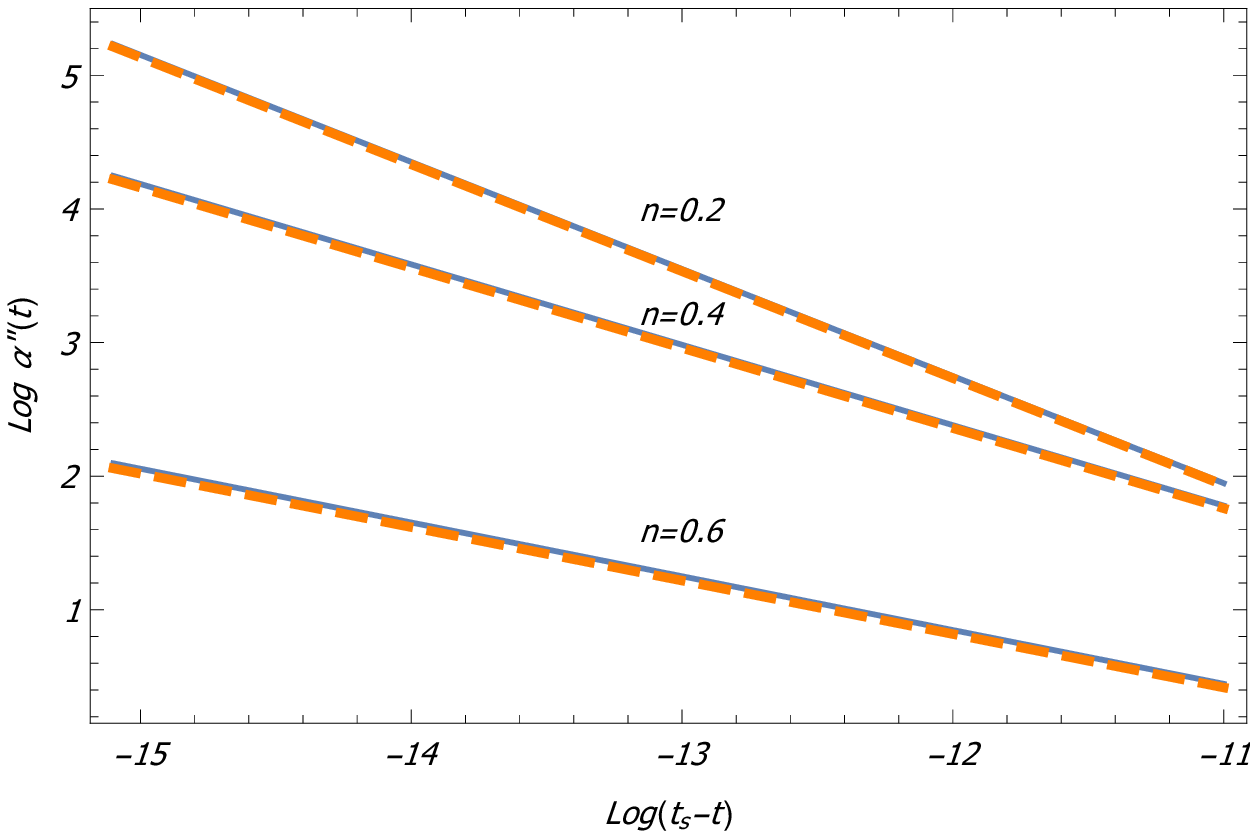}
\caption{4a}
\label{fig:fig5}
\end{subfigure}%
\begin{subfigure}{.5\textwidth}
\centering
\includegraphics[width=0.95\linewidth]{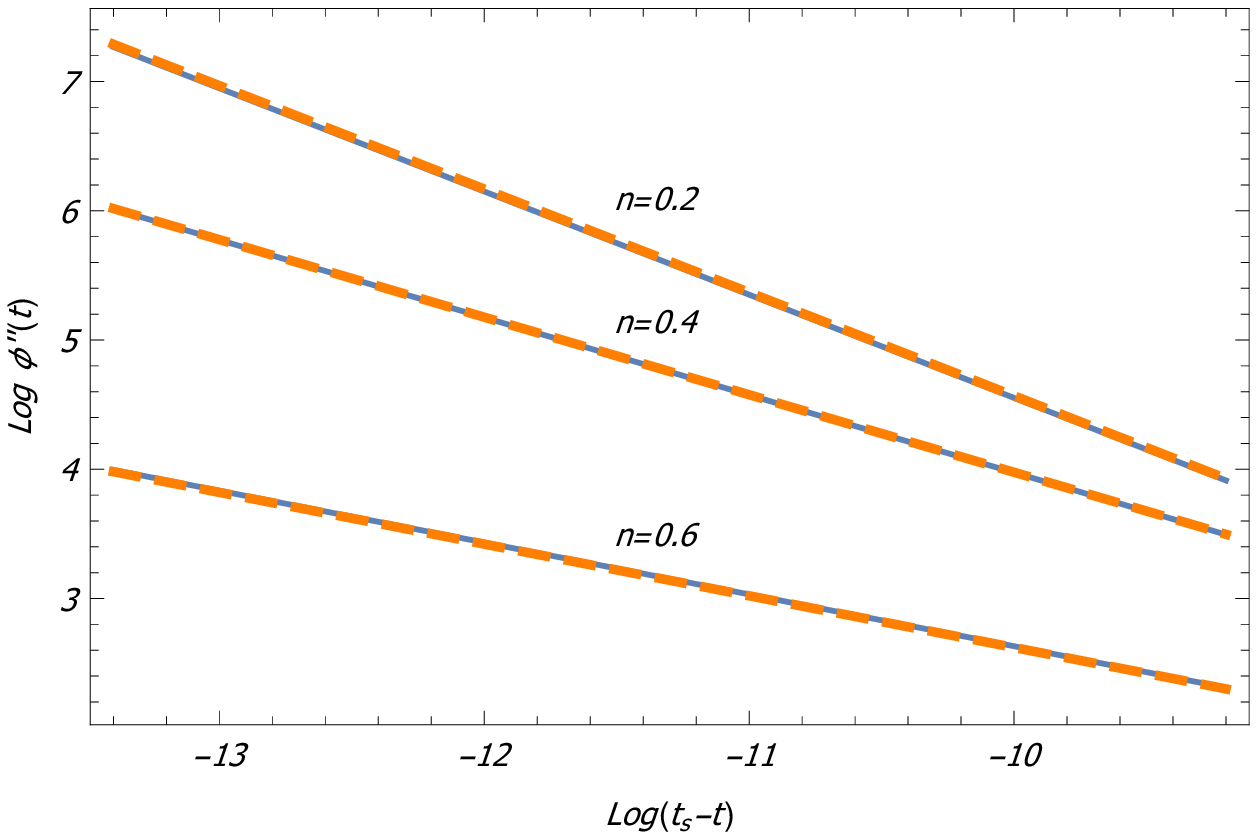}
\caption{4b}
\label{fig:fig6}
\end{subfigure}
\caption{Numerical verification of the $q$-exponent (4a) and $r$-exponent (4b), in the scalar-tensor case, for 3 values of $n$ ($n=0.2, n=0.4$ and $n=0.6$). The orange dashed line, denotes the analytical, while the blue line denotes the numerical solution. As expected the slopes for each n for both $q$ and $r$ are identical.}
\end{figure}

\begin{figure}[!h]
\centering
\begin{subfigure}{0.5\textwidth}
\includegraphics[width=0.95\linewidth]{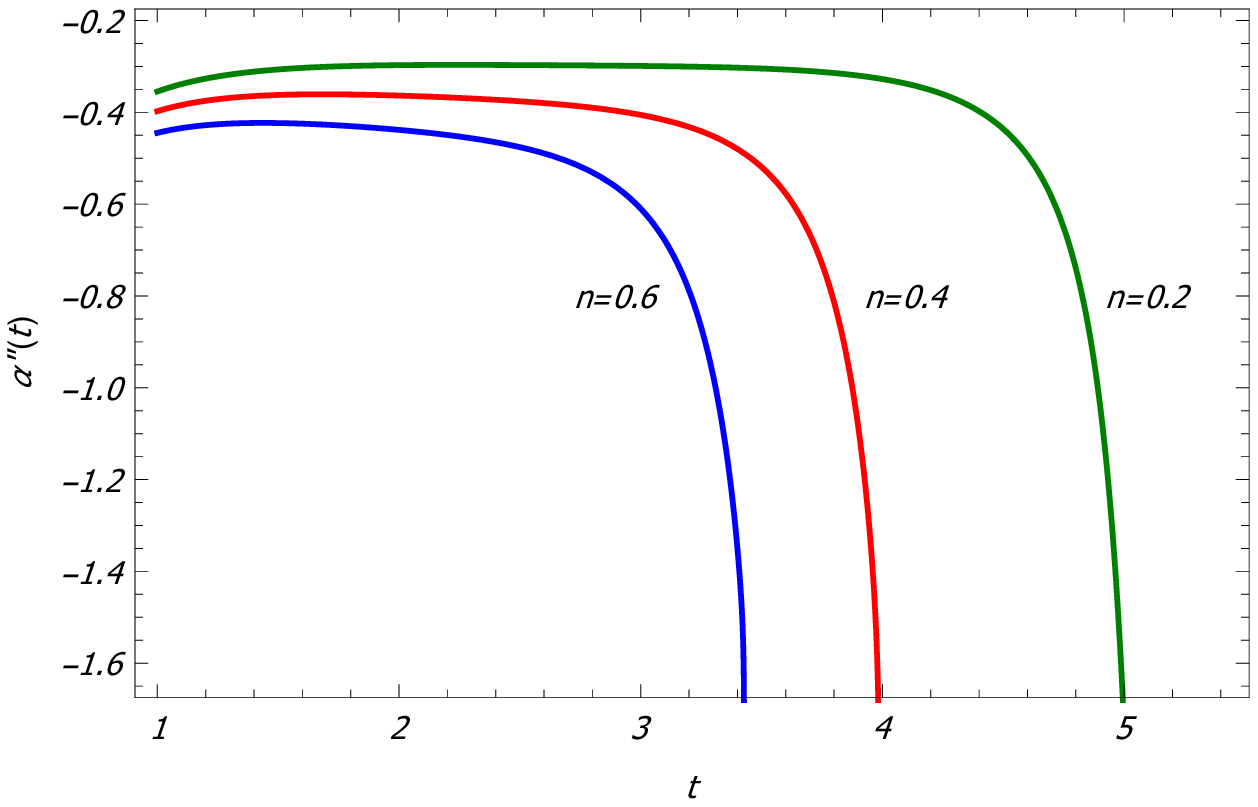}
\caption{5a}
\label{fig:fig7}
\end{subfigure}%
\begin{subfigure}{.5\textwidth}
\centering
\includegraphics[width=0.95\linewidth]{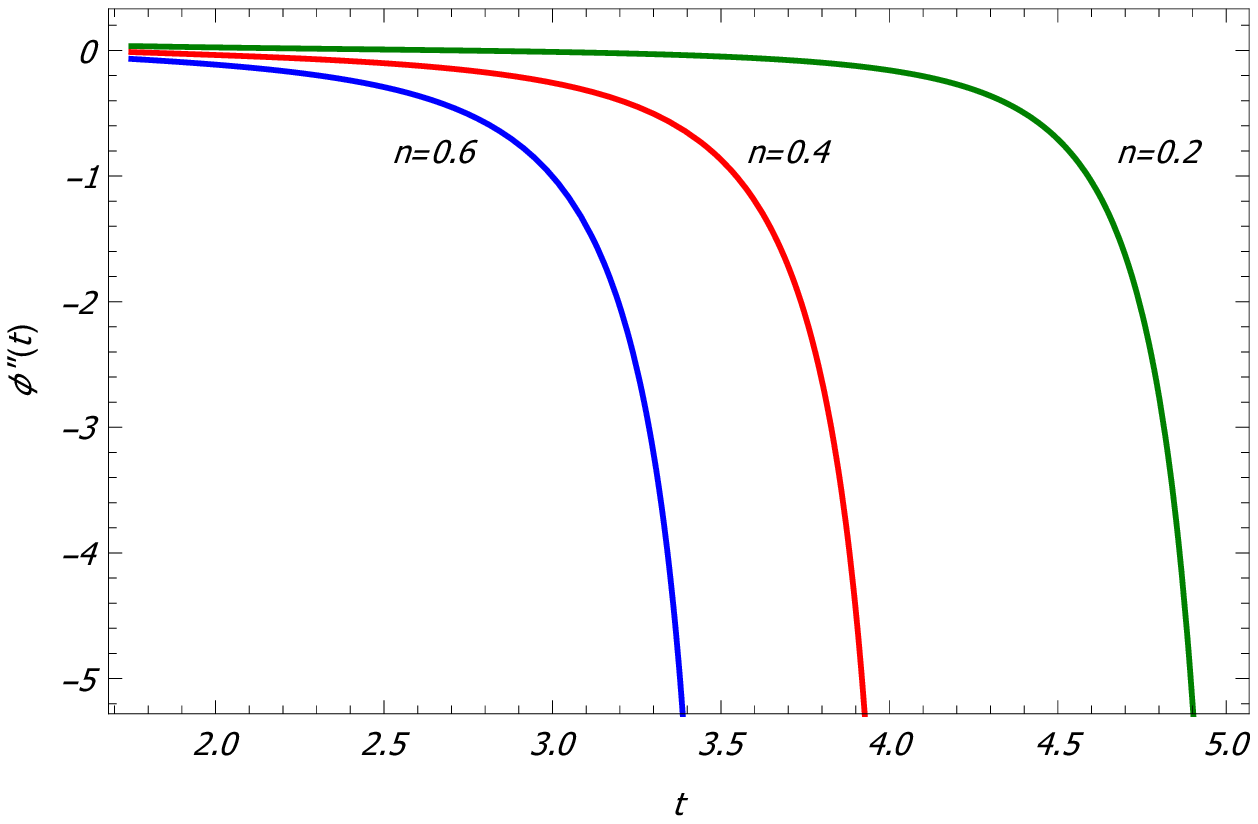}
\caption{5b}
\label{fig:fig8}
\end{subfigure}
\caption{Numerical solutions of the second time derivative of the scale factor (5a) and the scalar field (5b) for $n=0.2, 0.4, 0.6$. Notice the divergence of both the scale factor and scalar field at the time of the singularity.}
\end{figure}

In figures \ref{fig:fig5}, \ref{fig:fig6} we illustrate the numerically verified derived power law dependence eqs (\ref{str}), (\ref{stqn}) of the scalar field and the scale factor respectively, as the singularity is approached. Figures \ref{fig:fig7}, \ref{fig:fig8} depict the divergence of the second derivative, of both the scale factor and the scalar field, at the time of the singularity.

The results (\ref{str}) and (\ref{stqn}) are consistent with the above
qualitative discussion for the expected strength of the
singularity. Thus, in the case of the scalar-tensor theory, we have a
stronger singularity at $t_{s}$, as compared to the singularity that occurs in quintessence models. This is a general result, valid not only for the coupling constant of the form $F=1-\lambda \phi$ but also for other forms of $F(\phi)$ (\eg $F\sim \phi^r$), because the second derivative of $F$ with respect to time, in the dynamical equations, will always generate a second derivative of $\phi$ with divergence, leading to a divergence of $\ddot a$.

The quadratic term of $(t_{s}-t)$, in the expression of the scale factor (\ref{sfa}), is now subdominant as the second derivarive of the scale factor diverges. The only additional term of $(t_{s}-t)$ that can play an important role in the estimation of the Hubble parameter, is the linear term. Clearly, for the first derivative of (\ref{sfa}), as $t\to t_{s}$ from below, the linear term dominates over all other terms, while the quadratic term is subdominant in the second derivative, in the divergence of the $q$-term. Thus, in the case of the scalar-tensor quintessence models $H$ remain finite and dominated by the term $b(t_{s}-t)$, while $\dot H \to \infty$ as $t\to t_{s}$.

As in quintessence case of the previous section, in the absence of the perfect fluid, the strength of the singularity remains unaffected. This means that the evaluated relations of $r$ and $q$, eqs (\ref{str}), (\ref{stqn}) respectively, are exactly the same.

\section{Conclusions and Discussion}

We have derived analytically and numerically the cosmological solution close to a future-time singularity for both quintessence and scalar-tensor quintessence models. For quintessence, we have shown that there is a divergence of $\dddot a$ and a GSFS singularity occurs ($a_{s}, \rho_{s}, p_{s}$ remain finite but $\dot p \to \infty)$ , while in the case of scalar-tensor quintessence models there is a divergence of $\ddot a$ and an SFS singularity occurs ($a_{s}, \rho_{s}$ remain finite but $p_{s}\to \infty$, $\dot p \to \infty)$. In the absence of the perfect fluid in the dynamical equations, in both cases, we have shown that this result is still valid in our cosmological solution.

These are the simplest non-exotic physical models where GSFS and SFS singularities naturally arise. In the case of scalar-tensor quintessence models, there is a divergence of the scalar curvature $R=6\left (\frac{\ddot {a}}{a}+\frac{\dot a^{2}}{a^{2}} \right ) \to \infty$ because of the divergence of the second derivative of the scale factor. Thus, a stronger singularity occurs in this class of models. Such divergence of the scalar curvature is not present in the simple quintessence case.

We have also shown the important role of the additional linear and quadratic terms of $t_{s}-t$ in the form of the scale factor as $t\to t_{s}$. However, in the scalar-tensor case the quadratic term becomes subdominant close to the singularity.

For quintessence models, we derived relations of the Hubble parameter, $H^{2}(z)= \Omega_{0,m}(1+z)^{3}[1-(1+z)^{3}(1+z_{0})^{-3}]+(1+z)^{6}(1+z_{0})^{-6}H^{2}_{0}$ (for the fluid case) and $H(z)=\frac{H_{0}(1+z)^{3}}{(1+z_{0})^3}$ (for the no fluid case), close to the singularity. These relations may be used as observational signatures of such singularities in this class of models.

Interesting extensions of the present analysis include the study of the strength of these singularities in other modified gravity models \eg string-inspired gravity, Gauss-Bonnet gravity etc. and the search for signatures of such singularities in cosmological luminosity distance and angular diameter distance data. 

\section{Acknowledgments}

I would like to thank my collaborators S. Lola and L. Perivolaropoulos
for their stimulating and fruitful contribution and collaboration that
led to this work. I also thank the organizers for the opportunity to
present these results and for the stimulating atmosphere they have created during the conference. Financial support from the COST Action CA15108, is gratefully acknowledged.

\end{document}